\documentstyle[amssymb,aps]{revtex}

\begin{document}
\title{Primordial fluctuations from nonlinear couplings}
\author{Esteban A. Calzetta\thanks{%
email: calzetta@iafe.uba.ar} and Sonia Gonorazky\thanks{%
email: relat@iafe.uba.ar}}
\address{Instituto de Astronom\'\i a y F\'\i sica del Espacio (IAFE), c.c. 67, suc\\
28, (1428) Buenos Aires, Argentina and Departamento de F\'\i sica, Facultad\\
de Ciencias Exactas y Naturales, UBA, Ciudad Universitaria, (1428) Buenos\\
Aires, Argentina.}
\maketitle

\begin{abstract}
We study the spectrum of primordial fluctuations in theories where the
inflaton field is nonlinearly coupled to massless fields and/or to itself.
Conformally invariant theories generically predict a scale invariant
spectrum. Scales entering the theory through infrared divergences cause
logarithmic corrections to the spectrum, tilting it towards the blue. We
discuss in some detail whether these fluctuations are quantum or classical
in nature.
\end{abstract}

\pacs{98.80.Cq, 03.65.Sq, 04.62.+v}

\section{Introduction}

The inflationary scenario \cite{inflation} allows us to consider how the
primordial seeds of macroscopic structures were generated in the Universe,
since due to its quantum nature, the field that drives Inflation can be
decomposed into a mean field and fluctuations around it. The former gives an
homogeneous background of matter and the latter induce the production of
local inhomogeneities. These fluctuations evolve and are amplified during
the inflationary era. At the end of this epoch, the inflaton field decays
into relativistic ordinary matter. Heuristic arguments show that, as a first
approximation, a scale-invariant spectrum of density fluctuations results,
in rough agreement with observations \cite{observ}.

Despite this success, the conventional method of identifying the structure
creating fluctuations with the quantum fluctuations of the inflaton field
during slow roll is conceptually unsatisfactory, and eventually leads to an
overestimation of the produced density contrast (\cite{Mat}, \cite{CH95}) .
The basic point is that when we talk of the 'metric', or the density profile
of the Universe (as in 'a Friedmann - Robertson - Walker metric', or even
today, when we say 'space - time is flat' when talking about local physics)
we are referring to a macroscopic construct whereby microscopic (quantum)
fluctuations of geometry and matter fields are skipped over or coarse
grained away \cite{Hu95}. The difference between microscopic and macroscopic
fluctuations is not merely one of wavelength: the real difference is that
macroscopic fluctuations, when left to unfold over the relevant space and
time scales, effectively decohere from each other and thus acquire
individual reality. Indeed, this is the process by which a quantum
homogeneous state (such as the De Sitter invariant vacuum during slow roll)
may evolve into an inhomogeneous Universe: decoherence gives a formal
device, such as the harmonic analysis of quantum fluctuations, its physical
content. Now, given that macroscopic and microscopic fluctuations are to be
distinguished (and structure formation definitively belongs to the physics
of the former, as cosmic structures are 'classical', individually existing
objects), the relationship between them is not obvious and requires
elucidation. In the same way that the usual Brill-Hartle waves of general
relativity \cite{MTW}, being a first order effect, only react on the
background metric at second order, we should not expect microscopic
fluctuations by themselves to be lifted into the macroscopic level, but
rather that they will act on the macroscopic level as some higher order
effect. The goal of this paper is to present a detailed analysis of the
action of microscopic fluctuations on the macro level, obtaining from it an
improved estimate of the produced density contrast.

This issue can not even be posed correctly unless an open system view of the
inflaton dynamics is adopted. In this approach, the ''decoherence'', that
is, the conversion into c-number, of the q-number fluctuations is due to the
interaction of the inflaton field with a partially unknown and uncontrolled
environment. There are several proposals as to how the exact separation of
system and environment should be carried out (\cite{Mat}, \cite{PolStar}, 
\cite{MazLom}).

In this paper we shall present an improved discussion of to what extent
primordial fluctuations are ''quantum'' or ''classical'', from the viewpoint
of the ''consistent histories'' approach to quantum mechanics \cite{conhis}.
As it turns out, a detailed analysis of the conceptual difficulties of
Inflation points the way to the solution of the quantitative problems as
well (\cite{Mat}, \cite{CH95}, \cite{morikawa}).

The consistent histories approach views quantum evolution as the coherent
unfolding of individual histories for a given system, the main physical
input being the specification of the particular histories relevant to the
description of a concrete observer's experiences. For example, we could
choose our histories as containing an exhaustive description of the values
of all the fields in the theory at every space time location. A description
in terms of these "fine grained" histories is equivalent to a full quantum
field theoretic account of the dynamics. We shall rather assume that the
relevant histories for cosmological modeling are "coarse grained".

Concretely, we shall assume that close enough fine grained histories are
physically indistinguishable and should be bundled together as a single
coarse grained history. Each coarse grained history is thus labelled by the
value of a typical or representative history within the bundle. The actual
histories in a given bundle will differ from this representative by amounts
of the order of the quantum fluctuations of the corresponding fields (we
could consider also tighter or looser coarse graining, but as a matter of
fact these histories dominate the actual evolution of the system \cite{dch}).

Given a pair of coarse grained histories, we can compute the so-called
decoherence functional (df) between them. The df measures the quantum
overlap between these two histories. If the df between any two histories of
a given set is strongly suppressed, then quantum interference effects will
be unobservable, and it will be possible to treat each history classically,
that is, to assign individual probabilities to each of them. Moreover, the
most likely histories will be those for which the phase of the df is
stationary, which yields the ''equations of motion'' for the representative
history \cite{GelHar2}.

Going back to the problem of generation of fluctuations in Inflation, our
starting point is to assume that the evolution of the model is described in
terms of coarse grained histories as said, and to compute the df between two
generic coarse grained histories. We shall show that, for a variety of
models involving coupling the inflaton to massless fields of different spin,
coarse grained histories are indeed mutually consistent, and that the
equations of motion, as derived from the decoherence functional, are
stochastic. Thus, the representative fields naturally evolve fluctuations,
and these are responsible for the creation of primordial density
inhomogeneities at reheating.

It should be stressed that we are not assuming that the representative
fields are ''classical''; on the contrary, its classical nature is a
consequence of the theory itself, and follows from the suppression of the df
between generic coarse grained histories. Physically, the representative
field is decohered by its progressive entanglement with the microscopic
quantum fluctuations which surround it. This entanglement is a necessary
consequence of the nonlinear interaction between the two (for generic
initial conditions), and at the level of the equations of motion for the
representative field it appears as damping and noise. Thus, decoherence,
damping and noise are just different manifestations of the same process, a
point further elaborated elsewhere (\cite{CH95}, \cite{dch}).

In what follows, we shall consider inflationary models where the inflaton
field is nonlinearly coupled to itself, and to spin 1/2 and 1 massless
fields, respectively (the spin 2 case has been dealt with in ref. \cite{CH95}%
).

The paper is organized as follows. In next section, we consider in some
detail a simple model of Inflation, where the inflaton interacts with itself
through a cubic coupling. Treating the fluctuations around the
representative or physical value of the inflaton as a massless, minimally
coupled field, we shall derive the density contrast generated and discuss
both the amplitude of the scale invariant spectrum and the corrections to
it. In the following two sections, we briefly present the necessary
adaptations when the inflaton is coupled to massless, conformally invariant
spin 1/2 and 1 fields, respectively, and discuss the corresponding changes
in the predictions of the theory. We summarize our results in Section 5.

\section{Fluctuation generation from inflaton self - coupling}

\subsection{The model}

The production of the primordial seeds for structure generation began soon
after the set-up of Inflation and ended in the radiation dominated era.
Although realistic description of the phenomena that took place during this
epoch requires a detailed knowledge of the inflationary potential, it is
common to consider toy models that simplify the mathematical aspects of the
problem but are still accurate enough to give a qualitative description of
the related physics. We first consider a cubic field theory as a model for
the inflationary Universe 
\begin{equation}
V\left( \phi \right) =V(0)-\frac 16g\phi ^3  \label{pot}
\end{equation}
where $\phi $ is a c-number, homogeneous field, whose precise meaning shall
be discussed below. The dynamics of geometry is governed by the Friedmann
equation

\begin{equation}
H^2=\frac{V(\phi )}{m_p^2}
\end{equation}
where $H$ is the Hubble constant (we assume a spatially flat Friedmann -
Robertson - Walker (FRW) Universe and work, in this subsection, in the
cosmological time frame) and $m_p$ is Planck's mass. This equation assumes
vacuum dominance, namely

\begin{equation}
V(\phi )\gg \dot\phi^2
\end{equation}

We shall also assume potential flatness, that is

\begin{equation}
V(\phi )\sim V(0)\gg g\phi^3
\end{equation}

The field begins Inflation at some small positive value and then ''rolls
down'' the slope of the potential (at some point the potential must bend
upwards again, but that concerns the physics of reheating and shall not be
discussed here (\cite{reheat}, \cite{reh})). The dynamics of the homogenous
field is described by the Klein - Gordon equation

\begin{equation}
\ddot \phi +3H\dot \phi +(1/2)g\phi ^2=0
\end{equation}
(quantum corrections to this equation shall be discussed below). Under slow
roll over conditions ($\ddot \phi \ll 3H\dot \phi $) we find the solution

\begin{equation}
\phi (t)=\phi _0\left\{ 1-{\frac{g\phi _0t}{6H}}\right\} ^{-1}
\end{equation}

Slow roll over breaks down when 
\begin{equation}
1-\frac{g\phi _0t}{6H}\sim \frac{g\phi _0}{9H^2}
\end{equation}

Vacuum dominance applies to the whole slow rolling period under the mild
bound $\phi _0\leq m_p$. Potential flatness requires $H^4/g^2m_p^2\leq
10^{-3}$. Since $H$ is essentially constant during slow roll, the condition
for enough Inflation $Ht\geq 60$ implies $H^2\geq 10g\phi _0$. Current
bounds on $\Omega $ suggest that this bound is probably saturated; in this
regime the flatness condition is already satisfied given the other ones. The
final requirement on the model is enough reheating, namely $m_p^2H^2\leq
(T_{GUT})^4$.

The density contrast in the Universe is given in terms of the fluctuations
in $\phi $ by the formula \cite{GalForInf}. 
\begin{equation}
\left. \left( \frac{\delta \rho }\rho \right) _k\right| _{\text{in}}=\left. H%
\frac{\delta \phi _k}{\dot \phi }\right| _{\text{out}}  \label{hyperfam}
\end{equation}
which relates the density contrast at horizon entry to the amplitude of
fluctuations at horizon exit. Conventional accounts of the fluctuation
generation process estimate $\delta \phi _k$ from the value of the free
quantum fluctuations of a scalar field in a De Sitter Universe ($Hk^{-3/2}$
at horizon crossing) and thus find a Harrison - Zel'dovich (HZ) scale
invariant spectrum with amplitude 
\begin{equation}
{\frac{H^2}{\dot \phi }}\sim {\frac gH}\sim \sqrt{\frac g{\phi _0}}
\end{equation}
Thus, the observational bound of $10^{-6}$ on the density contrast implies $%
g\leq 10^{-12}\phi _0$.

One of the main aims of this paper is to present a different estimate. In
the approach to be presented below, the actual fluctuations in $\phi $ are
much less than expected (of order $gk^{-3/2}$), which leads to a revised
estimate $\delta \rho /\rho \sim (g/\phi _0)$ (no square root), and thus
relaxing the bounds on the self coupling by six orders of magnitude. This is
consistent with recent findings by Matacz and by Calzetta and Hu (\cite{Mat}%
, \cite{CH95}).

We proceed now to show how the revised estimate is found.

\subsection{Consistent histories account of fluctuation generation}

Let us now upgrade the inflaton field $\phi $ to a full fledged quantum
field $\Phi $ with a potential 
\begin{equation}
V\left( \Phi \right) =V(0)+c\Phi -\frac 16g\Phi ^3  \label{pot}
\end{equation}
(we have added the linear term for renormalization purposes). The massless
quantum field $\Phi $ obeys the Heisenberg equation of motion 
\begin{equation}
-\Box \Phi -\frac{dV}{d\Phi }=0  \label{ecmov}
\end{equation}

As described in the introduction, we shall assume that the fine details of
the evolution of the inflaton are inaccessible to cosmological observations.
Thus, we shall split the field as in

\[
\Phi =\phi +\varphi 
\]
where $\phi $ represents a typical field history within a bundle of
indistinguishable configurations, and $\varphi $ describes the unobserved
microscopic fluctuations. We identify $\phi $ with the classical inflaton
field of the previous subsection. $\varphi $ obeys linearized equations

\begin{equation}
-\Box \varphi -g\phi \varphi =0  \label{ecfluc}
\end{equation}

The equation for $\phi $ is obtained by substracting eqn. (\ref{ecfluc})
from eqn. (\ref{ecmov}) 
\[
-\Box \phi +c-\frac 12g\phi ^2-\frac 12g\left\langle \varphi ^2\right\rangle
_\phi =\frac 12g(\varphi ^2-\left\langle \varphi ^2\right\rangle _\phi ) 
\]
where $\left\langle ...\right\rangle _\phi $ means the expectation value of
the quantity between brakets, evaluated around a particular configuration $%
\phi $ of the physical field. If the constant $c$ takes the value $c=\frac 12%
g\left\langle \varphi ^2\right\rangle _0$ (which corresponds to evaluate $%
\left\langle \varphi ^2\right\rangle $ for the false-vacuum configuration $%
\phi =0$), and the right hand side is neglected, this equation admits the
false vacuum solution $\phi =0$ in a de Sitter geometry $g_{\mu \nu }=\frac 1%
{\left( H\tau \right) ^2}\eta _{\mu \nu }.$

We can also linearize this last expression to get the wave equation for
small fluctuations in $\phi $. The additional hypothesis that the phases of
the microscopic field $\varphi $ are aleatory assures that the right hand
side of this equation is always small. Indeed, if we were to identify $\phi $
with the expectation value of $\Phi $, we would drop this term altogether.
Since we are not doing such an identification, we shall retain it a little
longer, simply observing that we can evaluate this term at the false vacuum $%
\phi =0$ configuration:

\begin{equation}
-\Box \phi +\frac 12g\left( \left\langle \varphi ^2\right\rangle _\phi
-\left\langle \varphi ^2\right\rangle _0\right) =gj(\,x)  \label{ec26}
\end{equation}
where 
\begin{equation}
j(x)\equiv \frac 12\left[ \varphi ^2(x)-\left\langle \varphi ^2\right\rangle
_\phi (x)\right]  \label{fuente2}
\end{equation}
is seen as a noise source. The self correlation of this source is given by
the so called noise kernel (\cite{CH95}, \cite{CH94}). 
\begin{equation}
N(x_1,x_2)\equiv \frac 12\left\langle \{j(x_1),j(x_2)\}\right\rangle _\phi
\approx \frac 12\left\langle \{j(x_1),j(x_2)\}\right\rangle _0
\label{noiseker}
\end{equation}
The last term is a valid approximation provided the physical field $\phi $
remains close to its false vacuum configuration.

It is common to write eqn. (\ref{ec26}) as 
\begin{equation}
-\Box _x\phi \left( \,x\right) +g^2\int d^4x^{\prime }\sqrt{_{-}^{(4)}g}%
D\left( x,x^{\prime }\right) \phi \left( \,x^{\prime }\right) =gj(\,x)
\label{ec28}
\end{equation}
where 
\[
D\left( x,x^{\prime }\right) \equiv -\frac 1{2g}\frac{\delta \left\langle
\varphi ^2\right\rangle \left( \,x\right) }{\delta \phi \left( \,x^{\prime
}\right) }\mid _{\phi =0} 
\]
is the dissipation kernel (\cite{CH95}, \cite{CH94}). The physical meaning
of the noise and dissipation kernels is borne out by the df between two
histories described by different typical fields

\[
{\cal D}\left[ \phi ,\phi ^{\prime }\right] =\int D\varphi D\varphi ^{\prime
}\;e^{i\left( S\left[ \phi +\varphi \right] -S\left[ \phi ^{\prime }+\varphi
^{\prime }\right] \right) } 
\]
where the integral is over fluctuation fields matched on a constant time
surface in the far future. Actual evaluation yields (\cite{CH95}, \cite{CH94}%
)

\[
{\cal D}\left[ \phi ,\phi ^{\prime }\right] \sim \;e^{\left\{ iI-R\right\} } 
\]

\[
I=S\left[ \phi \right] -S\left[ \phi ^{\prime }\right] +(g^2/2)\int d^4x%
\sqrt{_{-}^{(4)}g}d^4x^{\prime }\sqrt{_{-}^{(4)}g^{\prime }}\left[ \phi
-\phi ^{\prime }\right] (x)D\left( x,x^{\prime }\right) \left[ \phi +\phi
^{\prime }\right] \left( \,x^{\prime }\right) 
\]

\[
R=(g^2/2)\int d^4x\sqrt{_{-}^{(4)}g}d^4x^{\prime }\sqrt{_{-}^{(4)}g^{\prime }%
}\left[ \phi -\phi ^{\prime }\right] (x)N\left( x,x^{\prime }\right) \left[
\phi -\phi ^{\prime }\right] \left( \,x^{\prime }\right) 
\]

We see that the dissipation kernel contributes to the phase of the df close
to the diagonal, and thus to the equations of motion for the most likely
histories, while the noise kernel directly determines whether interference
effects are suppressed or not, and thus the consistency of the chosen coarse
grained histories.

\subsection{Actual estimates of fluctuation generation}

The above treatment of fluctuation generation implies that there are
essentially two sources of fluctuations in $\phi $, namely, uncertainties in
the initial value data of $\phi $ at the beginning of Inflation, and
fluctuations induced by stochastic sources during the slow roll period (as
we shall see below, noise generation cuts off naturally after horizon
crossing).

Let us assume that decoherence is efficient (see below), and thus that we
can deal with each history individually. Then we must conclude that only
those histories where the initial value of $\phi $ is exceptionally smooth
may lead to Inflation (see Appendix). This limitation on initial data for
Inflation has been discussed by several authors, most notably from numerical
simulations by Goldwirth and Piran \cite{GoldPir}, and from general
arguments by Calzetta and Sakellariadou, Deruelle and Goldwirth \cite{infi}
and others. Discarding the fluctuations in the initial conditions, we find
the solution 
\[
\phi \left( \,x\right) =g\int d^4x_1\sqrt{_{-}^{(4)}g}G_{ret}(x,x_1)\,j(x_1) 
\]
where $G_{ret}$ is the scalar field retarded propagator, and the two-point
correlation function 
\[
\frac 12\left\langle \left\{ \phi \left( \vec x,\tau \right) ,\phi \left(
0,\tau \right) \right\} \right\rangle \approx g^2\int d^4x_1\sqrt{_{-}^{(4)}g%
}\int d^4x_2\sqrt{_{-}^{(4)}g}G_{ret}((\vec x,\tau ),x_1)G_{ret}((0,\tau
),x_2)N(x_1,x_2) 
\]

The noise and dissipation kernels can be written as 
\begin{equation}
N(x_1,x_2)\approx \text{Re}\left[ \left\langle j\left( x_1\right) j\left(
x_2\right) \right\rangle _0\right] =\text{Re}\left[ \left\langle \varphi
_1\varphi _2\right\rangle _0^{\,2}\right]  \label{njot}
\end{equation}

\begin{equation}
D(x_1,x_2)\approx \text{Im}\left[ \left\langle \varphi _1\varphi
_2\right\rangle _0^{\,2}\right] \,\theta \left( \tau _1-\tau _2\right)
\label{disip1}
\end{equation}

Returning to the fluctuation field associated to the $g\Phi ^3$ coupling, we
can write

\begin{equation}
\left\langle \varphi \left( x_1\right) \varphi \left( x_2\right)
\right\rangle _0=H^2\Lambda \left( r,\tau _1,\tau _2\right)  \label{capi2220}
\end{equation}
where $\Lambda $ is the dimensionless function: 
\begin{equation}
\Lambda \left( r,\tau _1,\tau _2\right) =\frac 1{2\pi ^2}\int_0^\infty \frac{%
dk}{k^2r}\sin \left( kr\right) f_k\left( \tau _1\right) f_k^{*}\left( \tau
_2\right)  \label{opapo}
\end{equation}
The $f_k$ are the positive frequency modes for the free-field in a de Sitter
geometry and are solutions for eqn. (\ref{ecfluc}) valid to first order.
These $f_k$ are functions of one single variable $k\tau _i.$

\[
f_k\left( \tau _i\right) =e^{ik\tau _i}\left( 1-ik\tau _i\right) 
\]
In the same 'first order' approach, we can consider that $G_{ret}$ is well
described by the free field retarded propagator: 
\[
G_{ret}(x,x_1)=-i\frac{H^2}{\left( 2\pi \right) ^3}\theta \left( \tau -\tau
_1\right) \int \frac{d^3k}{2k^3}e^{i\vec k\cdot (\vec x-\vec x_1)}\left\{
f_k(\tau )f_k^{*}(\tau _1)-f_k^{*}(\tau )f_k(\tau _1)\right\} 
\]

It is easy to see that the spatial Fourier transform of this quantity can be
written as $H^2k^{-3}{\cal G}\left( k\tau ,\beta _i\right) ,$ where we have
defined the dimensionless variable $\beta _i=k\tau _i$ and the wave-number
dependence has been factorized out of ${\cal G}$. Moreover, if we look at
eqns. (\ref{noiseker}), (\ref{capi2220}) and (\ref{opapo}) we conclude that
the spatial Fourier transform of the noise kernel can be written in
principle as $H^4k^{-3}{\cal N}\left( \beta _1,\beta _2\right) ,$ i.e. it
depends on $k$ only through the $k^{-3}$ factor. The Fourier transform of $%
\left\langle \phi \left( x_1\right) \phi \left( x_2\right) \right\rangle $
becomes

\begin{equation}
\Delta _k\left( k\tau \right) =\frac 14\frac{g^2}{k^3}\frac 1{\left( 2\pi
\right) ^6}\int_{-\infty }^{k\tau }\frac{d\beta _1}{\,\beta _1^4}%
\int_{-\infty }^{k\tau }\frac{d\beta _2}{\,\beta _2^4}{\cal G}\left( k\tau
,\beta _1\right) {\cal G}\left( k\tau ,\beta _2\right) {\cal N}(k,\beta
_1,\beta _2)  \label{ec18.1}
\end{equation}
The double-integral in eqn. (\ref{ec18.1}) represents a function of the
comoving wave number $k$ and the conformal time $\tau $ which appear in the
one-variable combination $k\tau .$

As we shall show below, fluctuation generation is effective only until
horizon crossing. As the $k$ mode of the field becomes greater than the
horizon when $k\tau =-1,$ the last consideration suggests that the integrals
in eqn. (\ref{ec18.1}) can be truncated at this value, and will therefore
take the form 
\begin{equation}
\Delta _k\left( k\tau =-1\right) =\frac 14\frac{g^2}{k^3}\frac 1{\left( 2\pi
\right) ^6}\int_{-\infty }^{-1}\frac{d\beta _1}{\,\beta _1^4}\int_{-\infty
}^{-1}\frac{d\beta _2}{\,\beta _2^4}{\cal G}\left( -1,\beta _1\right) {\cal G%
}\left( -1,\beta _2\right) {\cal N}(k,\beta _1,\beta _2)  \label{newver}
\end{equation}

If we take the above equations at face value, we find no explicit $k$
dependence within the integrand, and therefore the spectrum of field
fluctuations can be written as:

\begin{equation}
\Delta _k^{\text{sca}}\left( \tau \right) \propto g^2\frac 1{k^3}
\label{pesca5}
\end{equation}
where the superscript indicates that this prediction corresponds to a scalar
field theory. This is, of course, the well-established prediction of a
scale-invariant spectrum of density fluctuations.

However, in a de Sitter geometry a minimally coupled massless scalar field
is not well defined at the infrared limit, and the propagators associated to
it are divergent \cite{allen}. We can handle this problem by introducing an
infrared cut-off and studying the way in which this new parameter modifies
the Harrison-Zel'dovich spectrum (a small inflaton mass would have the same
physical effects). Our new propagator is: 
\begin{equation}
\Lambda _{\text{cut}}\left( r,\tau _1,\tau _2\right) =\frac 1{2\pi ^2}%
\int_{k_{\text{infra}}}^\infty \frac{dk}{k^2r}\sin \left( kr\right)
f_k\left( \tau _1\right) f_k^{*}\left( \tau _2\right)  \label{trunc}
\end{equation}

We want to find the $k$-dependence of $\Delta _k$ for the noise kernel
associated to the cubic coupling between two scalar fields, $\phi $ and $%
\varphi $. $N$ is obtained immediately from eqn. (\ref{njot}) as $N\left(
x_1,x_2\right) =H^4$ Re$\left[ \Lambda _{\text{cut}}^2\left( r,\tau _1,\tau
_2\right) \right] .$ As it was already noted, if we consider the retarded
propagators for the free field, then only ${\cal N}$ will have a non-trivial 
$k$-dependence. Of course, $k$ will always appear as an adimensional
quantity $k_{\text{infra}}/k$.

In order to analyze the emergence of corrective terms to a HZ spectrum, it
is convenient to note that ${\cal N}$ can be written as 
\[
{\cal N}={\cal N}_{\text{HZ}}+{\cal N}_{\text{infra}} 
\]
where ${\cal N}_{\text{HZ}}$ is independent of $k_{\text{infra}}$, and $%
{\cal N}_{\text{infra}}$ contains $k_{\text{infra}}$ only as $\ln \left(
k/k_{\text{infra}}\right) .$ It is now evident that, after performing the
double integration for the ${\cal N}_{\text{HZ}}$ term in eqn. (\ref{newver}%
), one will arrive to the usual $\Delta _k\propto k^{-3}$
Harrison-Zel'dovich spectrum. Furthermore, the logarithmic terms can be
factored out of the integral, i.e. the corrective terms to the spectrum will
have the form $\ln \left( k/k_{\text{infra}}\right) ,$ and eqn. (\ref{newver}%
) will just give the amplitude for this corrections. This means that,
provided the phenomena that induce the generation of fluctuations are
effective only until horizon-crossing and that it is a good approximation to
consider free-retarded propagators for the field, the spectrum of
fluctuations takes the form: 
\begin{equation}
\Delta _k^{\text{sca}}\left( k\tau =-1\right) =\frac C{k^3}\left[ 1+B\ln
\left( \frac k{k_{\text{infra}}}\right) \right]  \label{scapow}
\end{equation}
where $C$ is the Harrison-Zel'dovich amplitude and $B$ is the amplitude of
the corrections. An actual evaluation shows that $B$ is positive and that
its numerical value is about $5\times 10^{-3}$. This result tells us that
the logarithmic corrections increase the spectral power, specially for small
scales (large $k$), and the spectrum moves slightly to the blue.

As for the amplitude of the scale invariant part of the spectrum, we may
adopt the simple estimate eqn. (\ref{pesca5}). This leads to the revised
bound $g\leq 10^{-6}\phi _0$ discussed at the beginning of this section.

\subsection{Loose ends}

The method developed in this section to describe the generation of
primordial fluctuations can be applied with only trivial modifications to
other nonlinear theories involving the inflaton, as shall be demonstrated
below by considering couplings to spin 1/ 2 and 1 fields. However, before we
proceed, it is convenient to discuss in full two essential elements of our
argument, namely, that coarse grained histories described by generic values
of $\phi $ are truly consistent, and that super horizon fluctuations are
dynamically decoupled from the noise sources (more concretely, we must show
that on super horizon scales $\delta \phi _k\sim \delta \tau _k\dot \phi (t)$%
, since this formula enters the derivation of eqn. (\ref{hyperfam})).

Let us first consider the issue of consistency. We wonder if the history we
have considered, starting from vanishing initial conditions at the beginning
of Inflation, is truly decohered from any other history differing from it by
amounts of the order of the quantum fluctuations of an scalar field in De
Sitter space. If this is the case, then we are justified to treat this
history classically.

The answer to this question lies on whether the df between any such two
histories is strongly suppressed or not. In other terms, we must compute

\begin{equation}
-2\ln \{|{\cal D}[\phi ,\phi ^{\prime }]|\}\equiv g^2\int ~d^4x~\sqrt{-g(x)}%
~d^4x^{\prime }~\sqrt{-g(x^{\prime })}(\phi -\phi ^{\prime
})(x)N(x,x^{\prime })(\phi -\phi ^{\prime })(x^{\prime })
\end{equation}

Or, Fourier transforming on the space variables

\begin{equation}
g^2\int ~\frac{{d^3k}}{{(2\pi )^3}}~{\frac{d\tau }{(H\tau )^4}}~{\frac{d\tau
^{\prime }}{(H\tau ^{\prime })^4}}(\phi _k-\phi _k^{\prime })(t)~N_k(\tau
,\tau ^{\prime })(\phi _k-\phi _k^{\prime })(\tau ^{\prime })
\end{equation}

For each mode, the integral extends from the beginning of Inflation up to
horizon crossing. Due to the $\tau ^4$ suppression factor, the integral is
actually dominated by the upper limit. In this regime

\begin{equation}
N_k(\tau ,\tau ^{\prime })\sim H{^4/k^3}
\end{equation}

By choice, the value of the product of the fields is close to the
expectation value of quantum fluctuations, namely 
\begin{equation}
(\phi _k-\phi _k^{\prime })(\tau )(\phi _k-\phi _k^{\prime })(\tau ^{\prime
})\sim (H{^2/k^3)}\delta (0)\sim (H{^2/k^3)k_{{\rm infra}}^{-3}}
\end{equation}

(As follows from conventional quantization in the De Sitter background).

\begin{equation}
\int^{k^{-1}}~{\frac{d\tau }{(H\tau )^4}}\sim k{^3/}H{^4}
\end{equation}

Finally 
\begin{equation}
-\ln \{|{\cal D}[\phi ,\phi ^{\prime }]|\}\equiv \int \frac{{d^3k}}{{(2\pi
k)^3}}~\left( \frac gH\right) ^2({\frac k{k_{{\rm infra}}}})^3
\end{equation}

As we have seen, $g^2/H^2\sim g/\phi _0\sim 10^{-6}$, and so decoherence
obtains for all modes $k\gg 10^2k_{{\rm infra}}$. For example, if we take $%
k_{{\rm infra}}$ as corresponding to the horizon length at the beginning of
Inflation, and fine tune the model so that this will also correspond to the
horizon today, all modes entering the horizon prior to recombination would
be classical in this sense. Of course, in a realistic model $k_{{\rm infra}}$
would be much larger than today's horizon, and all physically meaningful
modes will be decohered. In this case, moreover, we would obtain decoherence
even between histories much closer to each other than the quantum limit.

Let us consider now the issue of noise on super horizon scales. In order to
arrive to the previous results, we have considered the integration of our
expression for the power spectrum of the fluctuations of the field (eqn. \ref
{ec18.1}) from the beginning of Inflation up to the moment in which each
mode $k$ crossed the horizon. The full expression can be rewritten as

\begin{eqnarray}
\Delta _k\left( \tau \right) =-\frac{g^2}{H^4}\frac 1{\left( 2\pi \right) ^6}%
\left\{ \int_{-\infty }^{-1}\frac{d\beta _1}{\beta _1^4}\int_{-\infty }^{-1}%
\frac{d\beta _2}{\beta _2^4}{\cal G}_1{\cal G}_2{\cal N}+\int_{-1}^{k\tau }%
\frac{d\beta _1}{\beta _1^4}\int_{-1}^{k\tau }\frac{d\beta _2}{\beta _2^4}%
{\cal G}_1{\cal G}_2{\cal N}+2\int_{-1}^{k\tau }\frac{d\beta _1}{\beta _1^4}%
\int_{-\infty }^{-1}\frac{d\beta _2}{\beta _2^4}{\cal G}_1{\cal G}_2{\cal N}%
\right\}  \label{gertru3}
\end{eqnarray}
where the second and third terms represent the contribution of a given mode
when it is outside the horizon. In the previous sections, we have ignored
these terms. If we consider the behavior of the noise kernel ${\cal N}$ far
away from the horizon, it is easy to verify that the last term may be
effectively ignored. The second term requires some additional
considerations. First we observe that the noise kernel is not oscillatory
outside the horizon, so the sources at different times are strongly
correlated. We can write $j_k\left( \tau \right) \sim j_k\sqrt{{\cal N}%
\left( k\tau \right) }$, where the $j_k$ are time-independent gaussian
variables. The wave equation that governs the evolution of each mode may be
written as 
\begin{equation}
-\ddot \phi _k+\left( H\tau \right) ^2k^2\phi _k\left( \tau \right)
+g^2\int_{-\infty }^\tau \frac{d\tau ^{\prime }}{\left( H\tau ^{\prime
}\right) ^4}{\cal D}\left( \tau -\tau ^{\prime }\right) \phi _k\left( \tau
^{\prime }\right) =gj_k\left( \tau \right) \simeq gj_k\sqrt{{\cal N}\left(
k\tau \right) }  \label{elena}
\end{equation}
where ${\cal D}$, $j_k$ and ${\cal N}$ indicate the spatial Fourier
transforms of the dissipation kernel, the source and the noise kernel,
respectively. When the mode is outside the horizon ($\left| k\tau \right|
\ll 1$) we can write the last equation as 
\begin{equation}
-\ddot \phi _k+g^2\int_{-\infty }^\tau \frac{d\tau ^{\prime }}{\left( H\tau
^{\prime }\right) ^4}{\cal D}\left( \tau -\tau ^{\prime }\right) \phi
_k\left( \tau ^{\prime }\right) \simeq gj_k\sqrt{{\cal N}\left( k\tau
\right) }  \label{elena217}
\end{equation}

The dissipative term is dominated by the contribution close to the upper
limit, and it can be written as: 
\begin{equation}
\ \frac{g^2}{H^4}\delta \phi _k\left( \tau \right) \int_{-\infty }^\tau 
\frac{d\tau ^{\prime }}{\tau ^{\prime \;4}}{\cal D}\left( \tau -\tau
^{\prime }\right)  \label{oop}
\end{equation}
The dissipation kernel can be obtained from eqn. (\ref{disip1}). The
asymptotic expressions near and far away from the coincidence limit $\tau
\simeq \tau ^{\prime }$ are, respectively:

\begin{eqnarray*}
\sqrt{-^{\left( 4\right) }g}{\cal D}_{\text{near}}\left( \tau -\tau ^{\prime
}\right) &\approx &\frac \pi {\tau ^{\prime \;4}}\left[ \frac 53\ln \left(
k\left( \tau -\tau ^{\prime }\right) \right) -1.50\right] \left( \tau -\tau
^{\prime }\right) ^3 \\
\sqrt{-^{\left( 4\right) }g}{\cal D}_{\text{away}}\left( \tau -\tau ^{\prime
}\right) &\approx &-\frac{\pi \tau }{2k}\frac 1{\tau ^{\prime \;3}}\sin
\left( k\left( \tau -\tau ^{\prime }\right) \right) \ln \left( k\left( \tau
-\tau ^{\prime }\right) \right)
\end{eqnarray*}
The upper formula holds when $\tau -\tau ^{\prime }\preceq k^{-1}$. ${\cal D}%
_{\text{near}}$ goes to zero rapidly as $\tau -\tau ^{\prime }\rightarrow 0$
and its contribution to the integral in eqn. (\ref{oop}) will be completely
negligible. Moreover, the oscillatory part of ${\cal D}_{\text{away}}$
cancels the contribution of the dissipative term far away from the
coincidence limit. From these observations, we conclude that dissipation is
not effective for modes that are outside the horizon, i.e. those modes
behave as a free field.

Since ${\cal N}\left( k\tau \right) $ grows at most logarithmically, we find
that the particular solution to eqn. (\ref{elena}) vanishes faster than $%
O\left( \tau \right) ,$ while the homogeneous (growing) solution is 'frozen'
into a constant value. Thus the value of $\phi _k$ obeys the usual
(classical) Klein - Gordon equation while beyond the horizon, and the
conventional derivation of eqn. (\ref{hyperfam}) holds \cite{GalForInf}.

\section{Yukawa coupling}

Now we consider the interaction between the inflaton field and a massless
Dirac field. The Lagrangian density for a theory in which two Dirac fields
are coupled to a scalar massless field is 
\[
{\cal L}=\partial _\mu \Phi \partial ^\mu \Phi +\frac i2\left[ \bar \Psi
\gamma ^\mu \partial _\mu \Psi -\partial _\mu \bar \Psi \gamma ^\mu \Psi
\right] +f\bar \Psi \Psi \Phi 
\]
where $f$ is an arbitrary coupling constant. The equation of motion for the
inflaton $\Phi $ is

\[
-\Box \Phi +m^2\Phi -f\bar \Psi \Psi =0 
\]
If we consider the separation of $\Phi $ in a mean field and fluctuations $%
\Phi =\phi +\varphi $ , the linearized equation of motion for the physical
field is 
\[
-\Box \phi \left( \,x\right) +m^2\phi \left( \,x\right) =f\,j_{\text{Yuk}%
}\left( x\right) 
\]
where 
\begin{eqnarray*}
j_{\text{Yuk}}\left( x\right) =\bar \Psi \left( x\right) \Psi \left( x\right)
\end{eqnarray*}
The noise kernel, defined as the mean value of the anticommutator of the
sources (see eqn. \ref{noiseker}), takes the form 
\begin{eqnarray}
N_{\text{Yuk}}\left( x_1,x_2\right) &\approx &\frac 12\left\langle \{j_{%
\text{Yuk}}(x_1),j_{\text{Yuk}}(x_2)\}\right\rangle _0  \label{yuk6} \\
&&  \nonumber \\
\ &\approx &\frac 12\left[ \left\langle \bar \Psi \left( x_1\right) \Psi
\left( x_1\right) \bar \Psi \left( x_2\right) \Psi \left( x_2\right)
\right\rangle _0+\left( 1\leftrightarrow 2\right) \right]  \nonumber
\end{eqnarray}

The four-point function can be reduced to a product of two-point functions
which correspond to the fermionic propagators 
\[
\left\langle \bar \Psi \left( x_1\right) \Psi \left( x_1\right) \bar \Psi
\left( x_2\right) \Psi \left( x_2\right) \right\rangle =\left\langle \bar 
\Psi \left( x_1\right) \Psi \left( x_2\right) \right\rangle \left\langle
\Psi \left( x_1\right) \bar \Psi \left( x_2\right) \right\rangle 
\]
where 
\begin{eqnarray*}
\left\langle \bar \Psi \left( x_1\right) \Psi \left( x_2\right)
\right\rangle &\equiv &-iS^{+}\left( x_2-x_1\right) \\
&& \\
\left\langle \Psi \left( x_1\right) \bar \Psi \left( x_2\right)
\right\rangle &\equiv &-iS^{-}\left( x_1-x_2\right)
\end{eqnarray*}

These expressions allow us to write the noise kernel as 
\begin{eqnarray}
N_{\text{Yuk}}\left( x_1,x_2\right) \approx -\frac 12f^{\,2}S^{-}\left(
x_1-x_2\right) S^{+}\left( x_2-x_1\right)  \label{nyuc321}
\end{eqnarray}
This expression is valid provided the scalar field remains near its false
vacuum configuration. As the spinor field is conformally invariant, the
propagators corresponding to a curved space-time can be written in terms of
those associated to a minkowskian geometry \cite{birrel}. For a de Sitter
background geometry, we have 
\[
S_{\text{dS}}^{\pm }\left( x_1,x_2\right) =H^3\left( \tau _1\tau _2\right)
^{3/2}S_{\text{Mink}}^{\pm }\left( x_1,x_2\right) 
\]
The minkowskian propagators for the spinor field can be written as
derivatives of the scalar field propagators 
\[
S^{\pm }=-i\gamma ^\mu \partial _\mu D^{\pm }=\pm \frac 1{\left( 2\pi
\right) ^3}\gamma ^\mu \partial _\mu \int \frac{d^3k}{2k}e^{i\left( \pm kx_0-%
\vec k\cdot \vec x\right) } 
\]

The noise kernel takes the form: 
\[
N_{\text{Yuk}}=-f^2H^4\left( \tau _1\tau _2\right) ^3\partial _\mu D_{\text{%
Mink}}^{-}\left( x_1-x_2\right) \partial ^\mu D_{\text{Mink}}^{-}\left(
x_1-x_2\right) 
\]

To arrive to a specific integral for the power spectrum generated by the
Yukawa coupling, we can proceed in close analogy to the scalar field case
(eqn. \ref{newver}):

\begin{eqnarray*}
\Delta _k^{\text{Yuk}}\left( k\tau =-1\right) &=&-\frac 14\frac{f^2}{k^3}%
\frac{H^4}{\left( 2\pi \right) ^6}\int_{-\infty }^{-1}\frac{d\beta _1}{%
\,\beta _1}\int_{-\infty }^{-1}\frac{d\beta _2}{\,\beta _2}{\cal G}\left(
-1,\beta _1\right) {\cal G}\left( -1,\beta _2\right) {\cal N}^{\text{Yuk}} \\
\ \qquad \qquad \qquad \qquad \qquad \qquad {\cal N}^{\text{Yuk}} &=&\frac 12%
{\cal F}\left[ \partial _\mu D_{\text{Mink}}^{-}\partial ^\mu D_{\text{Mink}%
}^{-}+c.c.\right]
\end{eqnarray*}
where ${\cal F}\left[ ...\right] $ represents the three-dimensional Fourier
transform of $\left[ ...\right] $.

As we are now considering conformal fields, the propagators are perfectly
defined and the last expression will produce a 'pure' HZ spectrum. As in the
scalar field, there are no relevant corrections coming from the ultraviolet
limit. The spectrum produced by this coupling will be of the scale invariant
form 
\[
\Delta _k^{\text{Yuk}}\left( k\tau =-1\right) =\frac{C^{\prime }}{k^3} 
\]
As a rough approximation, we may take ${C^{\prime }}\approx f^2H^4$, leading
to 
\[
\frac{\delta \rho }\rho \sim \frac{H\;\delta \phi }{\dot \phi }\sim \frac{%
H^3\;f}{g\;\phi ^2} 
\]
As we can write $H\sim \sqrt{g\phi }$ we obtain 
\[
\frac{\delta \rho }\rho \sim \sqrt{\frac g\phi }f 
\]
Given our previous estimate for the self-coupling, agreement between this
expression and the observational data requires that the coupling constant $%
f\sim 10^{-3}.$

\section{Electromagnetic coupling}

As a last example, let us consider the coupling between the inflaton field
and a massless vectorial field. The Lagrangian density for a theory with
massless scalar and electromagnetic fields is 
\[
{\cal L}=\left( \partial _\mu +ieA_\mu \right) \Phi \left( \partial _\mu
-ieA_\mu \right) \Phi ^{*}-\frac 14F^{\mu \nu }F_{\mu \nu } 
\]
from which we can deduce the equation of motion 
\[
-\left( \Box -e^2A^\mu A_\mu -ie\partial _\mu A^\mu \right) \Phi -2ieA^\mu
\partial _\mu \Phi =0 
\]
If we decompose the inflaton field in its physical and virtual components $%
\Phi =\phi +\varphi $ and write linearized equations, we obtain 
\[
-\Box \varphi =0 
\]
where we are assuming that $A^\mu $ is always small and can be thought as a
fluctuation in the electromagnetic potential $V^\mu =0.$ The more general
case $A^\mu =V^\mu +\delta A^\mu $ would give essentially the same results
for the small deviations $\delta A^\mu .$ The equation for the physical
field is: 
\[
-\left( \Box -e^2A^\mu A_\mu \right) \phi =2ieA^\mu \partial _\mu \varphi
+ie\left( \partial _\mu A^\mu \right) \varphi 
\]
The right hand side of this equation defines the source

\[
j\left( x\right) =\left[ A^\mu \partial _\mu \varphi +\frac 12\left(
\partial _\mu A^\mu \right) \varphi \right] 
\]

As usual, the noise kernel associated to this source is $N\left(
x_1,x_2\right) \simeq \left\langle \left\{ j\left( x_1\right) ,j\left(
x_2\right) \right\} \right\rangle _0$ where:

\begin{equation}
\left\langle j\left( x_1\right) j\left( x_2\right) \right\rangle _0=\qquad
\qquad \qquad \qquad \qquad \qquad \qquad \qquad \qquad \qquad \qquad \qquad
\qquad \qquad \qquad \qquad  \label{aa}
\end{equation}
\begin{eqnarray*}
&&\left\langle A^\mu \left( x_1\right) A^\nu \left( x_2\right) \right\rangle
_0\partial _{\mu ,1}\partial _{\nu ,2}\left\langle \varphi \left( x_1\right)
\varphi \left( x_2\right) \right\rangle _0+\frac 14\partial _{\mu
,1}\partial _{\nu ,2}\left\langle A^\mu \left( x_1\right) A^\nu \left(
x_2\right) \right\rangle _0\left\langle \varphi \left( x_1\right) \varphi
\left( x_2\right) \right\rangle _0  \label{camp147} \\
&& \\
&&+\frac 12\partial _{\nu _{,2}}\left\langle A^\mu \left( x_1\right) A^\nu
\left( x_2\right) \right\rangle _0\partial _{\mu ,1}\left\langle \varphi
\left( x_1\right) \varphi \left( x_2\right) \right\rangle _0+\frac 12%
\partial _{\mu ,1}\left\langle A^\mu \left( x_1\right) A^\nu \left(
x_2\right) \right\rangle _0\partial _{\nu ,2}\left\langle \varphi \left(
x_1\right) \varphi \left( x_2\right) \right\rangle _0
\end{eqnarray*}

Before we proceed, it will be convenient to write this expression in term of
the propagators of the interacting fields. The scalar propagator has been
considered in a previous section. It can be shown that a massless vectorial
field couples to the space-time curvature conformally. This result implies
that the covariant electromagnetic propagators for a de Sitter geometry are
identical to the minkowskian ones \cite{birrel}: 
\[
\left\langle A_\alpha \left( x_1\right) A_\beta \left( x_2\right)
\right\rangle _{\text{dS}}=\left\langle A_\alpha \left( x_1\right) A_\beta
\left( x_2\right) \right\rangle _{\text{Mink}}\equiv \left\langle A_\alpha
\left( x_1\right) A_\beta \left( x_2\right) \right\rangle 
\]
Rising indexes with $g^{\mu \nu }\left( x_i\right) =\left( H\tau _i\right)
^2\eta ^{\mu \nu }$ and adopting the Feynman gauge, where the minkowskian
electromagnetic and scalar propagators are related by 
\[
\left\langle A_\alpha \left( x_1\right) A_\beta \left( x_2\right)
\right\rangle =-i\eta _{\alpha \beta }D^{+}\left( x_1-x_2\right) 
\]
we obtain 
\[
\left\langle A_\alpha \left( x_1\right) A_\beta \left( x_2\right)
\right\rangle _{\text{dS}}=iH^4\left( \tau _1\tau _2\right) ^2\eta ^{\mu \nu
}D_{\text{Mink}}^{+}\left( x_1-x_2\right) 
\]

As this is a propagator for a conformal field, it is well defined and will
not produce any correction to the power spectrum. Only the factors which
correspond to the inflaton in eqn. (\ref{aa}) will produce corrections. In
order to get these corrections we must consider the truncated scalar
propagators defined in a previous section (see eqns. \ref{capi2220} and \ref
{trunc}). As we have already seen, the cut-off dependence can be isolated as 
$\log \left( \frac k{k_{\text{infra}}}\right) $, where $k$ is a parameter
that will be associated to the Fourier transform of the noise kernel. Thus
we can say that the sought for corrections will be logarithmic:

\[
\Delta _k^{\text{Em}}\left( k\tau =-1\right) =\frac{C^{\prime \prime }}{k^3}%
\left[ 1+B^{\prime \prime }\ln \left( \frac k{k_{\text{infra}}}\right)
\right] 
\]

As in the previous examples we considered, $C^{\prime \prime }$ is
undetermined because it includes the square of the coupling constant.
Roughly, $C^{\prime \prime }\sim H^4e^2,$ leading to $e\lesssim 10^{-3}$ to
match density production bounds. The ultraviolet contribution is always
irrelevant. $B^{\prime \prime }$ measures the relative importance of the
logarithmic corrections compared with the HZ background.

\section{Conclusions}

We considered fluctuation generation in the context of three elementary
regularizable field theories that represent the interaction of the inflaton
with itself and other massless fields of different spin. In each case, we
obtained the power spectrum for the field fluctuations $\Delta _k,$ which
can be easily related to the primordial density inhomogeneities that
constituted the seeds for structure generation. These fluctuations are
produced by a random noise source. We found that the predicted spectrum is
scale invariant when only conformal fields contribute to the noise term; in
a more general situation, such as when the source includes the virtual
scalar field, there appear logarithmic corrections.

Two features of our results stand out, namely, that we satisfy current
observational bounds on the amplitude of the primordial spectrum for values
of the inflaton self coupling much larger than previously reported, and that
the corrections to the HZ spectrum depend not only on the shape of the
inflaton potential, but also on what exactly the inflaton is coupled to.

Concerning the first issue, it should be clear that the drastic relaxation
on the bounds for the inflaton self coupling we have obtained is related to
much tighter bounds on the initial conditions for the inflaton field than
previously used. Of course, this is not the only factor that determines this
relaxation, for which we would also have to consider, at least, the {\it %
r\^ole} of the dissipation terms, which have been ignored so far. In this
sense, it might seem that we have just traded one fine tuning for another.
However, it should be remembered that the fine tuning of initial conditions
is not added ad hoc to match the COBE observations, but it is independently
necessary to obtain Inflation at all. As a matter of fact, this fine tuning
is necessary even if we accept the usual estimate of $g/\phi _0\sim 10^{-12}$%
. So, even if not yet totally satisfactory, it may be said that the model
has improved in regard to fine tuning. As we mentioned previously, a similar
result concerning fine tuning has been obtained by Matacz \cite{Mat} and by
Calzetta and Hu \cite{CH95}. Matacz considered a phenomenological model of
Inflation consisting of a system surrounded by an environment of time
dependent harmonic oscillators that back-react on the former acting as a
stochastic source of white noise. The approach by Calzetta and Hu consisted
on coarse-graining the graviton degrees of freedom associated to the
geometry of space-time. The latter methodology is followed closely in the
present work. We complement its results in some aspects such as making the
explicit calculation of the most relevant physical quantities, generalizing
the possible interactions of the scalar field and computing the main
corrections to the scale invariant spectrum.

In the long run, it may well be that the second aspect of our conclusions,
namely, the much wider scope to seek corrections to the fundamental Harrison
- Zel'dovich spectrum, will prove to be more relevant. Indeed, it is well
known that for any observed spectrum it is possible to ''taylor'' an
inflationary potential that will reproduce it \cite{taylor}. But these ad
hoc potentials have no other motivation that matching this result, and more
often than not are unmotivated or even pathological from the standpoint of
current high energy physics. The extra freedom afforded by the possibility
that the primordial spectrum of fluctuations could depend on the coupling of
the inflaton to other fields (which must exist if we are to have reheating)
could be the key to building simpler and yet more realistic theories of the
generation of primordial fluctuations.

Of course, the massless theories considered in this paper are too simplistic
to live up to this promise. Couplings to massive fields, and even the
possibility that the inflaton could be part of a larger, maybe grand
unified, theory, ought to be considered before actual predictions may be
extracted. We continue our research on this key issue in Early Universe
cosmology.

\section{Acknowledgments}

It is a pleasure to thank Antonio Campos, Jaume Garriga, Salman Habib,
Bei-lok Hu, Alejandra Kandus, Andrew Matacz, Diego Mazzitelli, Emil Mottola,
Juan Pablo Paz and Enric Verdaguer for multiple exchanges concerning this
project. We also wish to thank the hospitality of the Universidad de
Barcelona and the Workshop on Non Equilibrium Phase Transitions (Saint
John's College, Santa Fe, New Mexico, July 1996), where parts of it were
completed. This work has been partially supported by Universidad de Buenos
Aires, CONICET and Fundaci\'on Antorchas, and by the Commission of the
European Communities under Contract CI1*-CJ94-0004.\appendix 

\section{Homogeneous initial conditions}

In this appendix we try to clarify the reasons why we have considered
homogeneous initial conditions at the beginning of Inflation. We show that
the requirement of vacuum dominance at the beginning of Inflation excludes
classical fluctuations larger than $10^{-6}$ of the conventional vacuum
fluctuations on any interesting scale.

To start with, we consider the energy density associated to these classical
fluctuations

\begin{equation}
\rho \sim \dot \phi ^2+\nabla \phi ^2+V(\phi )
\end{equation}
as measured by a comoving observer. The fluctuations in the energy pick up
first and second order terms, which are denoted as $\delta \rho _1$ and $%
\delta \rho _2$ respectively: 
\begin{eqnarray}
\delta \rho _1 &\sim &\dot \phi _0\delta \dot \phi +V^{\prime }(\phi )\delta
\phi \\
\delta \rho _2 &\sim &\delta \dot \phi ^2+\nabla \delta \phi ^2+V^{\prime
\prime }(\phi _0)\delta \phi ^2.  \nonumber
\end{eqnarray}
We will only consider the second order terms. These terms dominate over $%
\delta \rho _1$ for small $\dot \phi _0.$ For modes far inside the horizon
the last term can be neglected. If the fluctuations behave as a massless
field, it follows that 
\begin{equation}
\delta \rho _k\sim \left( \delta \dot \phi _k\right) ^2=k_{phys}^2\left(
\delta \phi _k\right) ^2
\end{equation}
As usual, the scales $k$ refer to comoving quantities, while the $%
k_{phys}=k/a=k\left| H\eta \right| $ stand for quantities measured in terms
of physical lengths. The spatial average for the field fluctuations can be
written as 
\begin{equation}
\left\langle \delta \phi _k\,\delta \phi _{k^{\prime }}\right\rangle \approx 
\frac{\left( H\eta \right) ^2}k\sigma _k\delta (k-k^{\prime })
\end{equation}
where $\sigma _k$ measures the ratio between the classical fluctuations in
question and the quantum vacuum fluctuations in the De Sitter invariant
vacuum (which we include here only to have something familiar to compare
against). This expression allows us to write 
\begin{equation}
\left. {\frac{\delta \rho }\rho }\right| _{\text{in}}\sim \left. \left( 
\frac 1{m_pH}\right) ^2\int d^3k~\left( \frac ka\right) ^2\frac{\left( H\eta
\right) ^2}k\sigma _k\right| _{\text{out}}
\end{equation}
where we have used the Einstein equation to substitute $\rho \sim
(m_pH)^{-2},$ and we have assumed that the $\delta $-function in $%
\left\langle \delta \phi _k^2\right\rangle $ cancels the divergence
associated to the infinite volume over which this average is taken. If we
also assume that $\sigma _k$ obeys a power law, the last expression can be
written as 
\begin{equation}
{\frac{\delta \rho }\rho }\sim \left( \frac{k^4}{a^4m_p^{\,2}H^2}\right)
\sigma _k
\end{equation}
where $k$ corresponds to the lowest (in wavelenght) fluctuation scale. The
scale factor $a$ and the quantity $\sigma _k$ are evaluated at the beginning
of Inflation. In terms of the physical wavelength, we have 
\begin{equation}
{\frac{\delta \rho }\rho }\sim \left( \frac{l_p}{l_{hor}}\right) ^2\left( 
\frac{l_{hor}}{\lambda _{phys}}\right) ^4\sigma _k
\end{equation}
Obviously, the model is consistent only if this expectation value is lower
than one, i.e.: 
\begin{equation}
\sigma _k\le \left( \frac{l_p}{l_{hor}}\right) ^{-2}\left( \frac{l_{hor}}{%
\lambda _{phys}}\right) ^{-4}
\end{equation}
This means that we have an upper limit for the amount of classical
fluctuations which are present (at a given scale) at the beginning of
Inflation, if there is to be Inflation at all.

The question we must face now is whether this bound still allows for
fluctuations the size of those which will build up subsequently through
matching to an effective stochastic source, as we have demonstrated in the
main part of this paper. These latter fluctuations amount to around $10^{-6}$
of the quantum zero point fluctuations in the models we have considered.

The wavelengths where it is possible to have $\sigma _k\ge 10^{-6}$ (so that
initial classical fluctuations can dominate the fluctuations generated from
the stochastic source) must obey 
\begin{equation}
\lambda _{phys}\ge 10^{-3/2}\sqrt{l_p\times l_{hor}}
\end{equation}

We now recall the known result 
\begin{equation}
l_p\times l_{hor}=T_r^{\,-2}
\end{equation}
where $T_r$ is the reheating temperature. Thus, we have the condition 
\begin{equation}
\lambda _{phys}\ge 10^{-3/2}\frac 1{T_r}.
\end{equation}
It is convenient to phrase this condition in terms of the present wavelenght
of the same fluctuation 
\begin{equation}
\left. \lambda _{phys}\right| _{today}\ge 10^{-3/2}\frac 1{T_r}\times
e^N\times \frac{T_r}{T_0}
\end{equation}
where $N$ is the number of e-foldings during Inflation, $T_r$ the
temperature at reheating, and $T_0$ the temperature today.

In natural units, $T_0^{-1}\sim 10^{-28}d_0$, where $d_0$ is the present
size of the horizon. Thus, we must have 
\begin{equation}
\left. \lambda _{phys}\right| _{today}\ge 10^{-29.5}\times e^N\times d_0
\end{equation}
Most inflationary models predict values of $N$ over $60$ and even larger.
Thus, classical fluctuations that have $\sigma _k\ge 10^{-6}$ at the
beginning of Inflation are excluded on any cosmologically relevant scale.

\end{document}